\documentclass[final]{svjour2}
\usepackage{graphicx}
\usepackage{rotating}
\usepackage{amssymb}
\usepackage{mathptmx}
\usepackage[numbers]{natbib}
\makeatletter
\journalname{Journal of Low Temperature Physics}

\bibpunct{}{}{,}{s}{}{,}

\begin{document}

\def\he4{$^4$He}
\def\hee3{$^3$He}
\def\Am3{\AA$^{-3}$}
\def\beq{\begin{eqnarray}}
\def\eeq{\end{eqnarray}}
\newcommand{\hdblarrow}{H\makebox[0.9ex][l]{$\downdownarrows$}-}
\title{Interplay of non-linear elasticity and dislocation-induced superfluidity in solid \he4}

\author{D. Aleinikava and A.B. Kuklov}

\institute{Department of Engineering Science and Physics,
CSI, CUNY, \\ Staten Island, NY 10314, USA\\
Tel.:1-718-982-2887\\ Fax:1-718-982-2830\\
\email{Anatoly.Kuklov@csi.cuny.edu} }

\date{10.27.2011}

\maketitle


\begin{abstract}
The mechanism of the roughening induced partial depinning of gliding dislocations from \hee3 impurities is proposed as an alternative to the
standard "boiling off". We give a strong argument that \hee3 remains bound to dislocations even at large
temperatures due to very long equilibration times.
A scenario leading to the similarity between elastic and superfluid responses of solid \he4 is also discussed. Its main ingredient is
a strong suppression of the superfluidity along dislocation cores by dislocation kinks (D. Aleinikava, et. al., arXiv:0812.0983). 
These kinks, on one hand, determine the temperature and \hee3 dependencies of the generalized shear modulus  and, on the other, control the superfluid response. Several proposals for theoretical and experimental studies of solid \he4 are suggested.

PACS numbers: 
\end{abstract}

\section{Introduction}
The discovery of  the torsional oscillator (TO) anomaly by Kim \& Chan in 2004 \cite{KC}, which was originally ascribed to the non-classical moment of inertia (NCRI) \cite{Leggett}, has reignited a strong interest to the supersolidity \cite{SFS} in solid \he4. However, after observing the enigmatic similarities  between the TO and the shear modulus
by Day \& Beamish \cite{Beamish} (the similar effect has been observed by Tsymbalenko \cite{Tsymbalenko_84} at significantly higher
temperatures  and frequencies), serious doubts have been raised \cite{Reppy_2010} about the supersolid interpretation of the TO anomaly.
Very recently in Ref.\cite{Beamish_Balibar} it has been shown that the deformation of solid \he4 in the torsion rod is fully responsible for the anomaly reported by several groups.

A completely different approach has been undertaken by Ray \& Hallock \cite{Hallock} who have observed 
 the direct superflow through solid \he4 at very small rate -- about 3g/year. Thus, it is very likely that such a signal is below the sensitivity of the TO \cite{Hallock}. Under these circumstances, it is natural to ask: is there any "room" left for observing the supersolidity in the TO-approach? 

While not providing a definite  answer to this question, we show that the superfluid (SF) response of the dislocation cores (found to be SF in the {\it ab initio} simulations \cite{SFdisl,sclimb})
can mimic the temperature and \hee3 dependencies of the generalized shear modulus \cite{Beamish} (see Eq.(\ref{mu2})). The key to such a similarity is a possibility that dislocation kinks
strongly suppress the superfluidity along the core \cite{Aleinikava_2008} (cf. also a proposal by Balibar \cite{Balibar_2010}).      

Our paper is organized as follows: in Sec.\ref{sec_G} the shear modulus anomaly \cite{Beamish} is discussed in terms of the dislocation roughening in the presence of the Peierls and \hee3 pinning potentials. The inconsistency of the \hee3 "boiling off" scenario is revealed, and it is suggested that \hee3 remains bound to dislocations even at high $T$. The puzzle of the "missing dissipation" is addressed in Sec.\ref{diss}. Then, in Sec.\ref{SF} we will discuss the coarse grained model where the SF order parameter interacts with the thermal kinks, and will show that the condensate fraction can resemble the shear modulus $T$- and the \hee3- dependencies. We also discuss a  mechanism linking the dissipation in the  SF and in the gliding dislocation subsystems to the common bath of thermal kinks.
Finally, in Sec.\ref{sum} we give a brief summary of the results and outline proposals for future studies. 

\section{Generalized shear modulus}\label{sec_G}
The shear modulus anomaly of many materials is caused by liberation of gliding dislocations from the low temperature pinning by either impurities \cite{Friedel,Beamish}
or by the Peierls potential \cite{Aleinikava_2008,EPL}. As shown in Ref.\cite{Beamish_2010}, the reconciling of the observed $T$-dependencies of the shear modulus $G(T)$ with the "boiling off"-model of \hee3 impurities from the dislocation cores requires introducing a wide distribution of \hee3 activation energies $E_a$ -- as wide as the mean energy itself. The origin of such a wide distribution in a crystal with relatively low dislocation densities ( about $x_d \approx 10^{-7}-10^{-8}$ in the atomic units of the typical distance between \he4 atoms $b \approx 3.5-3.7$\AA) is very puzzling and raises more questions than gives answers. 

In this section we argue that the actual shape of  $G(T)$ in the presence of the Peierls potential and the \hee3 pinning
is determined by thermal roughening of gliding dislocations: thermal and quantum fluctuations of dislocation shape wash out both potentials. 
This process is the alternative to the "boiling off" mechanism. 

\subsection{Fluctuative depinning of gliding dislocation}\label{sub_model}
In solid \he4 dislocations are strongly pinned by the Frank's forest cross-linking points \cite{Friedel} and, in addition, are weakly pinned by
Peierls potential and \hee3 impurities. The model capturing both effects relies on the Granato-L\"ucke-type string \cite{Granato} description \cite{Aleinikava_2008,EPL}
subjected to both the Peierls potential $U_P$ and the trapping potential $U_t$ provided by \hee3 atom. The corresponding action capturing quantum and thermal effects can be written in the imaginary time $\tau$ as
\beq
H = \int_0^L dx \int_0^{\beta}d\tau \left( \frac{1}{2K}\left( (\nabla_xy)^2+(\nabla_\tau y)^2\right) - \sigma y - U_P(y)-N(x) U_t(y)\right),  
\label{model} \\
U_P(y)= - \alpha \cos ( 2 \pi y ), \quad  U_t(y)=N(x)V(y),\quad V(y)=-V_0 \exp ( - (y/y_0)^2 ) \label{UU},  
\eeq
where $y(x,\tau)$ is the dislocation displacement from its equilibrium position in a gliding plane; $L$ stands for the dislocation length between the cross-linking points ($y(x=0,\tau)=y(x=L,\tau)=0$); $\beta=1/T$ (in atomic units $\hbar=1, K_B=1$);  $K$ denotes the Luttinger parameter \cite{EPL}; $\sigma$ denotes external stress; $\alpha, V_0$ stand for the strength of the Peierls and the trapping potentials, respectively; $N(x)$ gives the density of \hee3 impurities which can be located at any point $0<x<L$ along the line $y=0$; the
trapping potential is chosen as a Gaussian with some range $y_0 \sim 1$.

In Eqs.(\ref{model},\ref{UU}) and below all lengths  are measured in units of Burger's vector $b$, and the unit of time is chosen so that the speed of sound $V_s$ along the core is unity. Accordingly, the unit of energy (temperature) is given by $T_o= \hbar V_s /K_B b =1$ (which is $\sim 10$K in standard units).  All our results below are represented in such dimensionless units.

The parameters can, in principle, be extracted from the {\it ab initio} simulations. For example, Ref.\cite{Corboz} provides the estimate $V_0 \approx 0.8$K for
binding of \hee3 to screw dislocation. In our simplified model we ignore the long-range part of the trapping potential, which should scale as  $V(y)\sim - 1/|y|$ for the case of edge dislocation (and as $\sim -1/y^2$ for the screw dislocation \cite{Corboz}). 

In Ref.\cite{Balatsky_2011} it has been suggested that \hee3 atoms don't actually provide pinning for dislocations, and, instead, they induce a viscous drag force.
Here we take a different approach: due to the extremely narrow band width $J\sim 10^{-4}$K \cite{Andreev_uspekhi}, a \hee3 atom-impuriton, once bound to a dislocation, can be easily localized in the lattice potential to become an immobile pinning center.

Monte Carlo simulations of  the partition function $Z= \\ \int Dy \exp(-H)$ as well as any needed mean $\langle ... \rangle = \int Dy ... \exp(-H)/Z$ have been conducted in discretized space-time lattice using the gradient expansion method combined with the Worm Algorithm \cite{WA} with the \hee3 impurities considered as Boltzmann particles of fixed total number, with $N(x)=0,1$ depending on if \hee3 atom is or is not present along a given space-time line $(x,\tau)$, respectively.   
This model is similar to the one considered in Ref.\cite{EPL} where $V_0=0$. Here we ignore the long-range potential between kinks \cite{EPL}.

The response of the model (\ref{model}) can be related to the average coarse-grained shear modulus \cite{EPL}. If $y(x,t)$ is a typical dislocation displacement of an element of Frank's forest of typical sizes $L, L_y, L_z$ (along the respective axes $x,y,z$), the average strain is $\approx \int y(x,t) dx/(LL_yL_z)$ (in the chosen units).  Since $y\propto \sigma$, and the total strain must include the elastic one $\sigma /G_{el}$, with $G_{el}$ being bare shear modulus, the resulting (inverse) shear modulus for static $\sigma$ becomes
$G^{-1}=G^{-1}_{el} + \Lambda \langle y \rangle/(LL_yL_z \sigma)$, where $\Lambda$ is some orientation averaging factor \cite{Granato} $\sim 1$.
In the limit $\sigma \to 0$ this gives
\beq
\frac{1}{G}=\frac{1}{G_{el}} + 
\frac{n_d\Lambda}{L} 
\int^\beta_0 d\tau \int_0^{L} dx \int_0^{L} dx' \langle y(x',\tau) y(x,0)\rangle
\label{G}
\end{eqnarray}
where $n_d= 1/(L_y L_z) \approx 1/L^2 <<1$ stands for density of dislocations in the units of $b$ and the mean $\langle ...\rangle$ is evaluated with respect to the action (\ref{model}) at $\sigma =0$.
\begin{figure}
\begin{center}
\includegraphics[%
  width=0.6\linewidth,
  keepaspectratio]{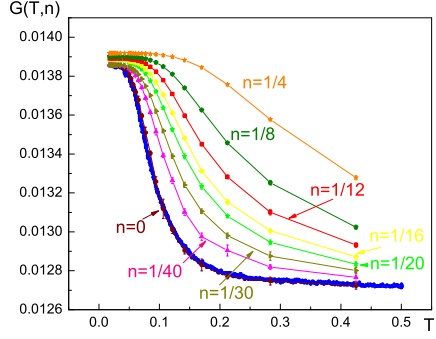}
\end{center}
\caption{(Color online) Shear modulus $G$ from Ref.\cite{Beamish} for 1 ppb of \hee3 at $f=2000$Hz (blue dots) and the MC results (\ref{model},\ref{G}) for various (linear) concentrations $n$ of \hee3 impurities. }
  \label{fig1}
\end{figure}
\begin{figure}
\begin{center}
\includegraphics[%
  width=0.6\linewidth,
  keepaspectratio]{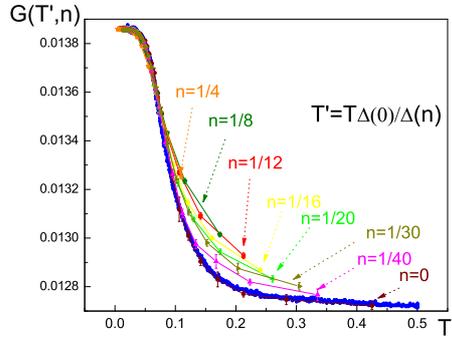}
\end{center}
\caption{(Color online) The experimental data \cite{Beamish} are the same as in Fig.~\ref{fig1}. The family of the MC curves from Fig.~\ref{fig1} has been collapsed to a single master curve by rescaling the $T$-axis and slightly adjusting $G_{el}$ to allow all curves to have the common value at $T=0$. The simulation parameters  are chosen as $K=0.1, V_0=0.3, y^{-1}_0=4.2$ and the Peierls potential amplitude $\alpha=0.01$ (in units of $T_o, b$) was adjusted to achieve best agreement with the experiment at $n=0$.  }
  \label{fig1_1}
\end{figure}
\begin{figure}
\begin{center}
\includegraphics[%
  width=0.75\linewidth,
  keepaspectratio]{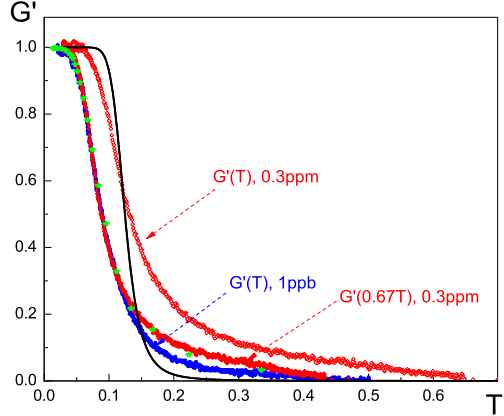}
\end{center}
\caption{(Color online) The shear modulus ($f=2000$Hz) normalized by its maximum variation  for two different \hee3 concentrations  1 ppb (blue dots) and 0.3ppm (open red rhombi) \cite{Beamish}.  
The data for 0.3ppm is also shown for the rescaled temperature $T \to 0.67T$ (filled rhombi). The Monte Carlo data (stars) for $n=1/40$ of \hee3 from Fig.~\ref{fig1} is  shown at the same rescaled temperature. Black solid line: the attempt to fit the 0.3ppm data by the "boiling off" model with the parameters chosen $n_d=0.3\cdot 10^{-8},\, E_a=0.65$K.}
  \label{fig3}
\end{figure}

The results of the simulations are presented in Figs.~\ref{fig1},\ref{fig1_1} for the case of pure Peierls potential ($n=0$, cf. \cite{EPL}) as well as for a set of finite linear densities $n$ of \hee3. For larger $n$ the softening of $G(T,n)$ occurs at larger $T$.
Such tendency is clearly seen  in Fig.\ref{fig1}. We have found that the curves for different $n$ are almost self-similar to each other and can be collapsed to a single master curve, Fig.~\ref{fig1_1}, by choosing a proper multiplicative factor $T_\Delta$: $T \to  T_\Delta T$. The simulations have been conducted for fixed values of $n$, that is, \hee3 atoms were not allowed to boil off from the dislocation. Despite that and the strong pinning at the dislocation ends,
thermal kinks induce fluctuations of the dislocation position $y$ around $y=0$ so that the Peierls and the \hee3-pinning potentials are essentially washed out, and the modulus can reach its high-$T$ value. 

The rescaling parameter $T_\Delta$ reflects the value of the dislocation gap $\Delta = T_\Delta^{-1}$ induced by the pinning potentials.
 We have found that  $\Delta$ grows with $n$. This can be understood from the following consideration: at small $n$ the effect of impurities is inducing some effective average potential $ \sim -nV_0\exp(-(y/y_0)^2) \approx -nV_0(1-(y/y_0)^2)$ in the action (\ref{model}). This corresponds to the formation of the effective gap $\sim \Delta \approx \sqrt{nV_0}/y_0$. As $n$ grows, the effect of the impurity-impurity repulsion through the string excitations becomes important (
the pinning increases the zero-point energy of the dislocation phonons and kinks). Thus, the gap must show a crossover from $\Delta \sim \sqrt{\Delta_0^2 + ...n}$ at small $n$ to   $\Delta \sim n$ 
at large $n$, where $\Delta^2_0 \sim \alpha$ is attributed to the Peierls potential. The fit by such quadratic function gives
$\Delta (n) = \sqrt{\Delta^2(0) + \delta_1 n +\delta_2 n^2}, \, \Delta(0)=0.209\pm 0.005,\, \delta_1=1.17 \pm 0.06,\, \delta_2=5.3 \pm 0.3$.

It is important that, while strongly fluctuating dislocation becomes practically free from the pinning potentials, it still provides a very strong binding potential $\tilde{V}_0$  to the impurities on average if compared to the impuriton band-width $J$. Indeed, we estimate $\tilde{V}_0$ as
$V(y)$, Eq.(\ref{UU}), smeared by large fluctuations of the dislocation position $y$ around $y=0$ as
\beq
\tilde{V}_0 \approx \frac{y_0}{\sqrt{\langle y^2 \rangle}}V_0 = y_0 \sqrt{\frac{T_D}{TL}}  V_0 << V_0,
\label{V0}
\eeq
where the thermal fluctuations are taken in the high-$T$ limit as $\langle y^2 \rangle \approx TL/T_D > y_0^2$ and $T_D$ stands for Debye temperature.
An estimate of $\tilde{V}_0/J$ for the typical values ($V_0 \sim 1$K,$T\sim 1$K, $T_D\sim 10$K, $y_0\sim 1$, $L\sim 1/\sqrt{n_d}\sim 10^4-10^5$)
shows that $\tilde{V}_0/J \sim 10^2$. As discussed below in Sec.\ref{Bof_he3}, such large ratio implies very long equilibration time for \hee3 to boil off into bulk.

\subsection{Fluctuative partial depinning versus boiling off of \hee3}\label{Bof_he3}
Figs.~\ref{fig1},\ref{fig1_1} represent the result of the fluctuative creep---quantum and thermal fluctuations of the dislocation shape due to kinks and phonons produce
decoupling of the dislocation line from the Peierls and \hee3-pinning potentials. In other words, despite the presence of impurities at the equilibrium dislocation position $\langle y \rangle=0$, the spatially averaged
pinning force acting on the dislocation is significantly diminished by the shape fluctuations. The question is how allowing \hee3 to boil off will modify
the curves $G(T,n)$. 

Let's, first, turn attention to the experimental situation. 
Fig.~\ref{fig3} shows Day \& Beamish data for $G(T)$ at two vastly different bulk concentrations of \hee3. These moduli can be collapsed on each other
and fit by the Monte Carlo data obtained under the assumption that \hee3 does not boil off. In contrast, the boiling off model with the single activation energy can fit neither of $G(T)$ (cf. \cite{Beamish_2010}). In other words, the experiment indicates that \hee3 remains bound to the dislocation even at high temperatures, which is consistent with the conjecture of very long thermalization time (as our estimates indicate --- much longer than any reasonable experimental time) of the impuritons. 

The equilibrium concentration $n(T)$ of \hee3 on dislocations at density $n_d$ 
can be found within the simplest thermodynamics consideration treating \hee3 atoms as non-interacting Boltzmann particles \cite{He3_us}:
\beq
n(T)=\frac{X_3}{n_d + \exp(-E_a/T)},
\label{n_trap}
\eeq
where we assume the total (bulk) fraction $X_3$ of \hee3   as $X_3 \leq n_d$ (in the chosen units) and $E_a$ stands for \hee3 activation energy.
This equation is also valid for $X_3>n_d$ as long as $T> E_a/|\ln(X_3-n_d)|$].

The boiling off starts at $T_{\rm bon} \approx E_a/|\ln(n_d)| << E_a$ for the realistic densities of dislocations $n_d \sim 10^{-6}-10^{-9}$. 
Given the typically accepted $E_a \sim 0.5-0.8$K, this gives the boiling off onset temperature as $T_{\rm bon} \sim 40-70$mK. The pinning becomes irrelevant when
the typical distance $r\sim 1/n$ between the trapped atoms is larger than $L \approx 1/\sqrt{n_d}$, Ref.\cite{Beamish}.
This determines the upper temperature $T_{\rm bup}$ when all \hee3 atoms have essentially evaporated as $T_{\rm bup}\approx E_a/|\ln(X_3/\sqrt{n_d})|$. For the standard
values of $X_3\sim 10^{-6}-10^{-9} <<1$ this gives $T_{\rm bup} \approx 70-100$mK implying that the actual softening range 
should be much narrower ($ \leq 30$mK) (see the solid black line in Fig.~\ref{fig3}) than it is observed experimentally \cite{Beamish}. 

The pure boiling off model corresponds to the action (\ref{model}) without Peierls potential ($\alpha=0$). In the approximation of strong pinning ($V_0 \to \infty$)
by the impurities placed (equidistantly a distance $r=1/n(T)$, Eq.(\ref{n_trap}), apart from each other) along the dislocation of the length $L\approx 1/\sqrt{n_d}$  one finds (using the free string solution \cite{Granato} between two pinning centers) the shear modulus (\ref{G}) 
\beq
\frac{1}{G(T)}=\frac{1}{G_{el}} + \frac{\gamma}{1+ n^2/n_d}, \quad \gamma \equiv \frac{1}{G(\infty)} - \frac{1}{G_{el}}.
\label{strong2} 
\eeq
This equation  together with Eq.(\ref{n_trap}) have been used to obtain the solid fitting curve in Fig.~\ref{fig3} for the parameters typical for solid \he4.
As can be seen, this fit is not adequate \cite{note}. 

Similarly, the attempt to include the boiling off mechanism into the Monte Carlo data by replacing $n$ by the form (\ref{n_trap}) leads to the same failure if one
chooses the realistic values of $n_d, E_a, X_3$. Simply saying, if the equilibrium were reached fast on the experimental scale, there would be no impurities left on dislocations to affect the curve $G(T)$ at $T \geq 0.07-0.1$K. Nevertheless, as discussed above, the fluctuative mechanism completely ignoring the  evaporation of \hee3 from dislocations can describe well the temperature dependence of $G(T)$ in a wide range of \hee3 concentrations.  In other words, it appears that the actual equilibrium between binding to dislocations and being free in the bulk cannot be established for \hee3 atoms in any reasonable experimental time. Accordingly, the amount of \hee3 trapped by a dislocation is predetermined during
crystal growth.     

The following considerations may help to understand our conjecture: \hee3 atoms in a solid \he4 exist as impuritons 
with a very narrow band-width $J$ so that any repulsive or attractive potential characterized by energies larger than $J$ causes 
a \hee3 atom to be localized within a spatial region of a size on which the potential energy changes by not more than $J$, Ref.\cite{Andreev_uspekhi}. 
Furthermore, bulk phonons which are responsible for establishing the equilibrium can only change an impuriton energy by $J$ which is much smaller than $\tilde{V}_0$, Eq.(\ref{V0}).
In other words, there is no chance for an impuriton to be freed into the bulk by just absorbing the necessary thermal
energy during one scattering event. Instead, it must diffuse through the energy landscape with the help of many scattering events $\sim (\tilde{V}_0/J)^2 >>1$. 
Since the corresponding scattering matrix elements are suppressed as high powers of $T/T_D$ and $J/T_D$, Ref.\cite{Andreev_uspekhi}, such diffusion must be strongly
suppressed.    
The detailed solution of this problem along the line of the approach \cite{Kagan} will be presented elsewhere.

The inconsistency of the boiling off mechanism has already been noted in Ref.\cite{Beamish_2010}.
Here we have shown that the alternative scenario is the fluctuative partial depinning of dislocation from the Peierls and \hee3 pinning potentials.
Within this scenario the corresponding activation energy seen experimentally \cite{Iwasa,Palanen,Beamish_2010} should be attributed to the kink-pair creation energy 
(dependent on strengths of the both potentials) rather than to the activation of \hee3 from a stationary dislocation.

\section{The puzzle of "missing dissipation"}\label{diss}
Both phenomena---the NCRI and the shear anomalies---are characterized by the dissipation peak in the region of fast variation of  $G(T)$ and the
NCRI fraction $\rho_s(T)$. 
The single-time relaxation scenario for such peak appears to be inadequate \cite{Dorsey} because 
the relaxation induced change of the real part appears to be significantly larger than it should follow from the observed maximum
of the imaginary part of the generalized response. This situation can be referred to as the puzzle of "missing dissipation".
In order to fix the problem on empirical level, various  wide distributions of relaxation times can be invoked \cite{Balatsky, Beamish_2010}. 
 As we have mentioned above, such conjectures, while providing a convenient empirical framework,
do not explain the origin of the wide distribution in relatively good quality crystals. 

Here we argue that, in addition to the dynamical variations of the real and imaginary parts of the response which are connected
by the Kramers-Kronig relation, Peierls and \hee3 pinning potentials induce a purely equilibrium temperature-variation of the real part (at zero frequency $\omega$)
\cite{EPL} which is not directly linked to any dissipative process. This relaxes the "constraint" between the real and imaginary parts of $G$.

In order to illustrate the two origins of the shear modulus softening we introduce a linearized version of the model (\ref{model}) which takes into account the pinning gap $\Delta$: all the non-linear terms are replaced by $ \sim K^{-1}\Delta^2 y^2/2$.
The corresponding  dynamical equation of the string biased by a small external stress $\sigma(t)\sim \exp(-{\rm i} \omega t)$
\beq
\ddot{y}(x,t) - \nabla_x^2 y(x,t) + \int_{0}^\infty dt' \epsilon(t') y(x,t-t') +  \Delta^2 y(x,t) =\sigma(t),
\label{eq_dis}
\eeq
with the boundary condition $y(x=\pm L/2,t)=0$, in addition to the standard relaxation-dissipation term $\sim \epsilon$ (see in Ref.\cite{Granato}), includes the gap term $\sim \Delta^2$
induced by Peierls and \hee3 pinning. As shown above, $\Delta$ can be found from purely thermodynamical simulations of the full model (\ref{model}) (see also in Ref.\cite{EPL}).

The nature of the dissipative kernel $\epsilon(t')$ at low $T$ represents a separate fundamental problem.
In this regard, we note the proposal \cite{Markelov} that dislocation kinks initiate strong dissipation in solid \he4 \cite{dis}.
Similarly, the origin of the so called Bordoni peak in the internal friction in 
metals has also been assigned to kinks \cite{Bordoni}.  Both approaches rely on the standard treatments in terms of the kinetic
equation, without, however, addressing explicitly the issue of the integrability of the Sine-Gordon
model (that is, that there should be no dissipation in it!).

Here we conjecture that the main source of the dissipation are the kink-kink collision processes described in Ref.\cite{Sachdev} under the assumption of the integrability broken by the spatial discretization of the Sine-Gordon description. 
Then, the long-time kink motion is described by diffusion coefficient $D \sim \exp(\Delta/T) /\Delta$, Ref.\cite{Sachdev}. This corresponds to $  \epsilon(t') \sim \delta(t') /D = C_o \Delta \exp(-\Delta/T)$ in Eq.(\ref{eq_dis}) (where $C_o \sim 1$ in chosen units), so that the space-dispersive relaxation is given by $ \sim D q^2$ with $q$ standing for wave-vector along the dislocation. Such form  implies that the dissipation vanishes as $T\to 0$, when the kink density $ \sim \exp(-\Delta/T) \to 0$, and also as the gap $ \Delta (T) \to 0$ at large $T$, when kinks overlap and loose their meaning.  

Now we calculate the generalized shear modulus $G(\omega)$ in the frequency domain as $
G^{-1} = G^{-1}_{el} + (\Lambda n_d /L) \int^L_0 (y/\sigma) dx$ (see the explanation above Eq.(\ref{G})). Then solving Eq.(\ref{eq_dis}) in Fourier with the zero-boundary condition we obtain
\beq
G^{-1}(T,\omega)&=&G_{el}^{-1} + \frac{\Lambda_0 }{[\Omega (T,\omega) L/2]^2} \left(1- \frac{\tanh[\Omega(T,\omega)L/2]}{\Omega(T,\omega)L/2}\right),
\label{G_G} \\
\Omega^2(T,\omega)&\equiv& -{\rm i}\omega C_o \Delta {\rm e}^{-\Delta/T} - \omega^2 + \Delta^2.
\label{Gam}
\eeq  
The constant $\Lambda_0\equiv \Lambda n_d L^2$ (which is independent of $L$!) is determined by the total softening  effect. Specifically, at $T=0$ and $\omega =0$, $\Omega\sim {\cal O}(1)$ and $ G^{-1} - G_{el}^{-1} \sim 1/L^2 \to 0$ for $ L>> 1$, and at $\omega=0$ and large $T$ when $ \Delta =0$: $ G^{-1}(\infty,0) - G_{el}^{-1}= \Lambda_0/12$. 

\begin{figure}
\begin{center}
\includegraphics[%
  width=0.65\linewidth,
  keepaspectratio]{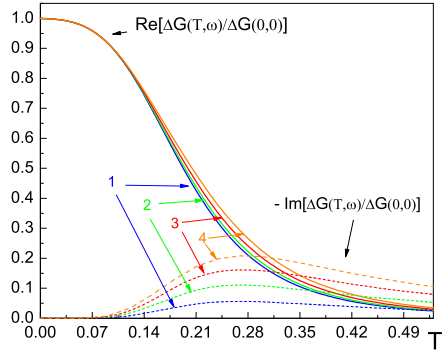}
\end{center}
\caption{(Color online) The real (solid lines) and imaginary (dashed lines) parts of the variation of the generalized shear modulus $G$ given in Eqs.(\ref{G_G},\ref{Gam}) normalized by the total variation of $G$ from $T=0$ to $ T= \infty$. The family of four curves corresponds to four values of the dissipation strength $C_o$ shown in the plot. The other chosen parameters are: $\Delta_1=0.1,\, \Delta_0 =0.05,\, L=100, \omega =0.003, \, \Lambda_0 G_{el}=0.4$ (in the dimensionless units discussed in the main text).} \label{Re_Im}
\end{figure}

A particular form of $\Delta(T)$ fitting well the experimental data was found to be $\Delta(T) = \Delta_1 (1-\exp(-\Delta_0/T))^2$, where $\Delta_0$ depends
on the strength of the Peierls and  \hee3 pinning potentials as well as on $n$ and $\Delta_1 \sim 0.1$ is a constant. We have used these expressions in Eqs.(\ref{G_G},\ref{Gam}) and have calculated the real and imaginary parts of the shear modulus, Fig.~\ref{Re_Im}.
As can be seen, in contrast to the simplistic single-relaxation-time model \cite{Dorsey,Beamish_2010}, the variations of the real and imaginary parts are not restricted by the condition that the maximum of the imaginary part is given by one-half of the total variation of the real part. Instead, while the real part is weakly dependent 
on the dissipation strength $C_o$ and is mostly controlled by the $T$-dependence of the gap $\Delta$, the imaginary part is strongly dependent on the value of the friction coefficient $C_o$, and, in principle,
can be zero.

\section{Kink-controlled core superfluidity}\label{SF}
Here we will give details of the proposal \cite{Aleinikava_2008} linking the shear modulus anomaly to the superfluid response.
Its main ingredient is a strong interaction between the core superfluidity \cite{SFdisl,sclimb} and density of kinks along the core.
We will take into account this effect within the coarse grained mean field description relying on the bulk-averaged SF order parameter
$\psi$ as well as on the bulk-averaged density of kinks $n_k$. 
The interaction free energy between the two is taken in the minimal form $F_{\rm int} = g_1 n_k |\psi|^2 >0$,
with $g_1>0$ being some phenomenological coefficient. Thus, the full bulk free energy density takes the form
\beq
F= a'(T-T_0) |\psi|^2 + \frac{g_2}{2}|\psi|^4 + g_1 n_k |\psi|^2 
+\Delta(T) \cdot n_k + Tn_k\ln(n_k/(n_0e)),  
\label{F_L}
\eeq
where $a'>0, g_2>0$ are the standard Landau-expansion coefficients with $T_0$ corresponding to the
mean field SF-transition temperature. The term $\sim \Delta$ describes energy of thermal activation of kinks
and the last one accounts for the entropy of the kink gas in the limit $n_k << n_0$, where $n_0$ stands for the maximum bulk
concentration of kinks determined by the size of the Sine-Gordon soliton  $ \propto 1/\sqrt{\Delta}$ and the density of dislocations $n_d$. Here kinks are treated
as Boltzmann particles. 

We note that, formally, the model (\ref{F_L}) apart from the kink-terms coincides with the approach \cite{PWA_science}.
However, in contrast to Ref.\cite{PWA_science} advocating a true supersolidity in \he4, we view the formation of the SF
order parameter as being due to a network of dislocations with superfluid cores \cite{Shev,SFdisl,sclimb}. In this analysis we don't consider
that some dislocations exhibit superclimb  and the anomalous compressibility \cite{sclimb}. The coarse grained description of such
network is a separate interesting problem.
 
\subsection{The Mean Field phase diagram}
The equilibrium of the system can be determined from the minimization $\delta F/\delta \psi^*=0$ and $\delta F/\delta n_k=0$:
\begin{eqnarray}
[a'(T-T_0) + g_1n_k]\psi+ g_2 |\psi|^2\psi=0, \quad
n_k=n_0\exp\left({-\frac{\Delta + g_1|\psi|^2}{T}}\right). 
\label{n}
\end{eqnarray}
The first equation can be formally written as a relative value of the condensate density $\rho_s(T)=|\psi|^2$ as
\begin{eqnarray}
\tilde{\rho}(T)=\frac{\rho_s(T)}{\rho_s(0)}= 1- \frac{T}{T_0} - \frac{g_1n_k}{g_2\rho_s(0)}, \,\,\, 
\label{til_rho} 
\end{eqnarray}
where $\rho_s(0)=a'T_0/g_2$ and we took into account that $n_k(T=0)=0$. 

\begin{figure}
\centerline{\includegraphics[angle = 0,
width=0.55\columnwidth]{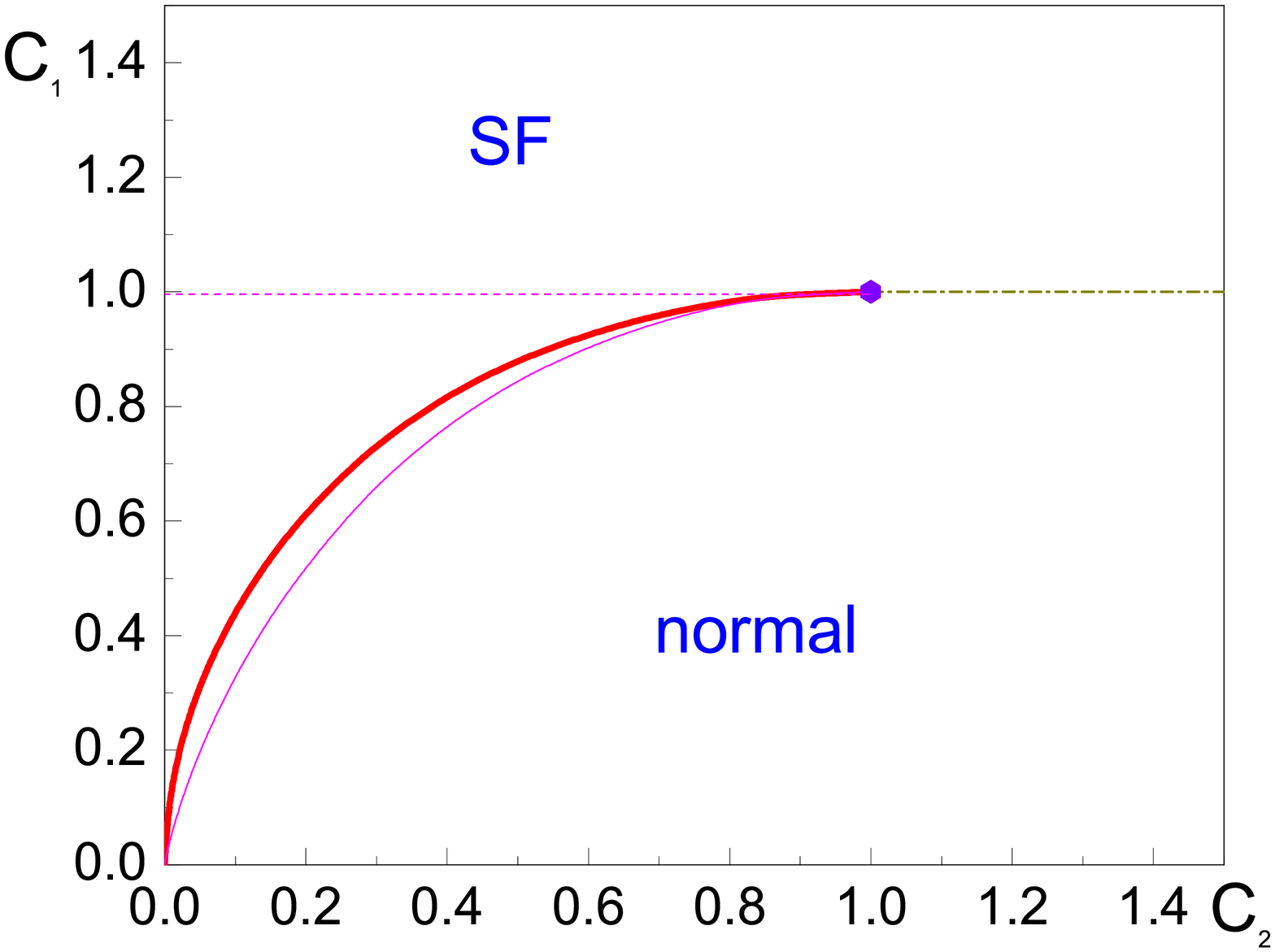}} \caption{(Color online)
The mean field phase diagram of the model (\ref{n}) in terms of the variables (\ref{par}). Solid red and dashed dark yellow lines represent the I and II order phase transitions, respectively. The large blue dot denotes the tricritical point separating the II and I order transitions. The domains above and below the transition lines marked as SF and as "normal" correspond to the SF and the normal states, respectively. The thin pink curve indicates the spinodal $C_s(C_2)$, Eq.(\ref{spinodal}). The metastable superfluidity can occur above this line and below the pink dashed straight line $C_1=1$, $C_2<1$.
} \label{PD}
\end{figure}

Eqs.(\ref{n}) describe the mean field phase diagram featuring lines of II and I order transitions with the tricritical point separating them.
It is convenient to introduce the following dimensionless quantities 
\begin{eqnarray}
C_1=\frac{a'(T_0-T)}{g_1n_0}\exp\left({\frac{\Delta}{T}}\right)>0 ,\quad
C_2=\frac{g_2T}{g_1^2n_0}\exp\left({\frac{\Delta}{T}}\right)>0,  
\label{par}
\end{eqnarray}
for $T<T_0$. Then, the II order (mean field) transition takes place along the line $C_1=1, C_2>1$. 
Along this line $F$ has only one minimum $\psi=0$ for $C_1<1$ and $\psi \neq 0$ for $C_1>1$, with $n_k \neq 0$ in both phases.
As $C_2<1$, $F$ can have two coexisting minima. That is, the I order transition occurs for $C_2<1$, 
 with the tricritical point  located at $C_1=C_2=1$. 
The actual equation for the I order transition line (when both minima have equal free energies) cannot be solved explicitly. 
Its numerical solution is represented by thick red line in Fig.~\ref{PD}.

The boundaries (spinodals) within which $F$ has two minima, that is, where metastable superfluidity can exist, can be found analytically.
These are given by $C_2<1, C_s(C_2)<C_1<1$ where 
\begin{eqnarray}
C_s(C_2)=C_2\ln\left(\frac{e}{C_2}\right),\,\, 0<C_2\leq 1.
\label{spinodal}
\end{eqnarray}
  
\subsection{Similarity between static superfluid and mechanical responses}\label{simil}

Here we will discuss the striking similarity between the SF response $\rho_s(T)$ and $G(T)$ following
from the model introduced above (cf. in Ref.~\cite{Beamish}).
\begin{figure}
\centerline{\includegraphics[angle = 0,
width=0.55\columnwidth]{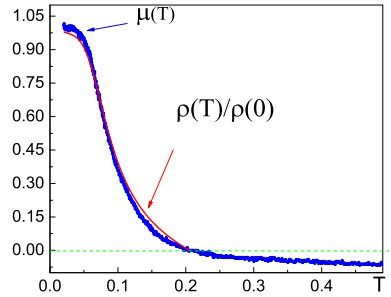}} \caption{(Color online)
Blue dots: Relative shear modulus, Eq.(\ref{mu_T}), with $T_c=0.21K$, from ~\cite{Beamish}. Red line -- relative superfluid density obtained from Eq.(\ref{til_rho}). Green dashed line shows zero for $\rho_s(T)$ at $T=T_c$.
} \label{sim}
\end{figure}
It occurs along the II order line due to the strong interaction between kinks and superfluidity given by the  term $\sim g_1$ in Eq.(\ref{F_L}). 
Let's consider the system far from the tricritical point, $C_2>>1$, and close to the II order line, $C_1\approx 1$. The condition $C_1=1$ determines the mean field transition temperature $T_c$ as
\begin{eqnarray}
\exp\left(\frac{\Delta}{T_c}\right) (1-T_c/T_0) = \frac{g_1n_0}{a'T_0}.
\label{C1}
\end{eqnarray}
If 
\begin{eqnarray}
\frac{g_1n_0}{a'T_0}>>1, 
\label{strong}
\end{eqnarray}
which constitutes the condition of strong kink-superfluid interaction, one finds
\begin{eqnarray}
T_c \approx \frac{\Delta}{\ln(g_1n_0/a'T_0)} << T_0.
\label{Tc}
\end{eqnarray}
At this point it is important to mention that the temperature $T_c$ is rather a renormalized value of $T_0$ than the actual transition of establishing the 3D coherence. In other words, it is the {\it onset} temperature where the NCRI should appear in a sense of the Shevchenko state ~\cite{Shev} (cf. the vortex-fluid model \cite{PWA}).

The condition that the system is far from the I order transition $C_2 >>1$ in combination with $T_c <<T_0$, Eq.(\ref{strong}), gives 
\begin{eqnarray}
\frac{gT_c}{g_1a'T_0}>>1 . 
\label{C2}
\end{eqnarray}
To what extent this condition is satisfied in solid \he4 remains to be seen. 

Let us demonstrate the similarity between the mechanical and superfluid responses in the limit (\ref{strong}). In the limit (\ref{C2}) $T$ should be dropped from the definition of $C_1$ in Eq.(\ref{par}) for $0<T<T_c$, so that 
the solution for the superfluid fraction $\rho_s(T)=|\psi|^2$, Eqs.(\ref{n}),  can be formally rewritten in the form
\begin{eqnarray}
\frac{\rho_s(T)}{\rho_s(0)}= 1- \frac{n_k(T)}{n_k(T_c)}, \,\,\, T\leq T_c
\label{rhos}
\end{eqnarray}
where $n_k(T)$ obeys Eq.(\ref{n}) and $T_c$ is given in Eq.(\ref{Tc}). 

The shear modulus thermal softening discussed above in Sec.\ref{sec_G} is determined by the averaged density of thermal kinks $n_k$. 
Indeed, Eq.(\ref{G}) relates the $G(T)$ softening to the fluctuations of the string displacement $\langle y^2 \rangle$ which, in its turn, is determined
by the density of thermal kinks $n_k$. Qualitatively, one finds  $\langle y^2 \rangle \propto n_k$ at least at low $T$ where kinks are still well defined entities. 
Thus, the shear modulus $G$ given by Eq.(\ref{G}) can be rewritten in terms of the coarse grained kink density as
\begin{eqnarray}
G(T)=\frac{G_0}{1 + \tilde{A} n_k(T)},
\label{G_n}
\end{eqnarray}
where $G_0 = G(T=0) \approx G_{el}$ and the prefactor 
$\tilde{A}$ can be taken as a constant of $T$ for all practical purposes because it does not contain any exponential. 

We introduce the relative change of the shear modulus with respect to its value $G(T_c)$ at the transition temperature (where $\rho_s$ is zero):
\begin{eqnarray}
\mu(T)= \frac{G(T) - G(T_c)}{G(0) - G(T_c)}.
\label{mu_T}
\end{eqnarray}
Substituting $G$ from Eq.(\ref{G_n}), we find
\begin{eqnarray}
\mu(T)= \frac{1}{1+\tilde{A}n_k(T)}\left(1- \frac{n_k(T)}{n_k(T_c)}\right)\approx \frac{\rho_s(T)}{\rho_s(0)}
\label{mu2}
\end{eqnarray}
where the last line follows from Eq.(\ref{rhos}) and the factor $1/(1+\tilde{A}n(T)) \approx 1$  because it controls the total relative variation of the modulus within only 5-15\% ~\cite{Beamish}. Thus, within such accuracy, our model naturally features the close similarity between the relative variations of the superfluid and mechanical responses. 

It should, however, be mentioned that the above simplified model cannot be applied to temperatures significantly above the gap value $\Delta(T=0)$, that is, when the simple activation dependence $n_k \propto \exp(-\Delta/T)$ does not fit well (see in Ref.\cite{EPL}) the actual experimental data Ref.\cite{Beamish}. Nevertheless,  Eq.(\ref{til_rho}) does not actually rely on the simple activation form of the kink density and will remain the same regardless of a particular form of the kink free energy. Thus, the actual form of $n_k$ at elevated $T$
can be extracted from the experiment \cite{Beamish} with the help of Eq.(\ref{G_n}). Then, substituting the so found $n_k(T)$ into Eq.(\ref{til_rho}) and using $g_1/a'$ as the only adjustable parameter for chosen $T_0=1K$, we find $T_c=0.21K$ and obtain the graph Fig.\ref{sim}. It shows the data from Ref.\cite{Beamish}, $f=2000$Hz, plotted together with the relative superfluid density $\tilde{\rho}_s(T)$ obtained from Eq.(\ref{til_rho}). The similarity between the two responses is obvious.

Concluding this section, we note that the energy $\Delta$ sets the scale for both---$G(T)$, below which it stiffens, and for the onset temperature $T_c$ of the SF response. As discussed in Sec.\ref{sub_model}, adding \hee3 increases $\Delta$ and, therefore, increases the SF onset temperature.

\subsection{Similarity between the dissipative responses.}

The similarity between the SF and mechanical responses considered above is not limited
to the static case.  Obviously, a strong coupling $\sim g_1$  must induce the same dissipation in both subsystems.  
As a result, excitations in the SF network will show damping similar to the one introduced into Eq.(\ref{eq_dis}).
Such excitations lead to a dynamical mass (and the moment of inertia) redistribution inside the TO cell. Accordingly, the TO response should
show the dissipation as long as SF currents percolate along a non-uniform network which, in general, allows linear coupling between SF phase and rotation. 
We demonstrate the similarity between the dissipation (in the shear modulus and in the SF excitations)  in the mean field equations for small oscillations of
the SF order parameter. Using the hydrodynamic approximation $\psi=\sqrt{\rho} {\rm e}^{i \phi}$ for the conjugate variables $\rho, \phi$ in $F$, Eq.(\ref{F_L}),
and the canonical commutation relations, we find Hamiltonian equation for $\rho$ from (\ref{F_L}). For $n_k$ as classical particles we use linearized kinetic equation (cf. \cite{Markelov}) in the single-time $\tau$ relaxation approximation which ignores spatial dispersion: $\dot{n}_k = - \frac{1}{\tau} \delta F /\delta n_k$. This gives 
\beq
\ddot{\rho} - \rho_0 \vec{\nabla}^2 (g_2 \rho + g_1 n_k)=0, \quad \dot{n}_k = -\frac{1}{\tau}(n_k - n^{(eq)}),
\label{rho}
\eeq  
where $\rho_0$ denotes the coarse grained SF density; $n^{(eq)}$ stands for the quasi-equilibrium density of kinks given by Eq.(\ref{n}).
The small variation $\delta n^{(eq)}$ of $n^{(eq)}$ induced by the time dependent part $\rho'$ of $\rho=\rho_0 + \rho' $ can be found from minimizing the free energy (\ref{F_L}) in the linear approximation as:
\beq
\delta n^{(eq)}= - \frac{g_1 n^{(eq)}_0}{T} \rho', \quad n^{(eq)}_0=n_0\exp\left({-\frac{\Delta + g_1\rho_0}{T}}\right)
\label{neq}
\eeq
where $n^{(eq)}_0$ stands for the static value of the kink density found from the second equation in (\ref{n}).
A substitution of $\delta n^{(eq)}$ into Eqs.(\ref{rho}) and further elimination of $n_k$ in Fourier gives the dispersion relation for the SF excitations
\beq
-\omega^2 + c_0^2 q^2\left[1- \epsilon'_0 \frac{{\rm i}\omega \tau}{1 -{\rm i}\omega \tau}\right]=0,\quad \epsilon'_0\equiv \frac{g_1^2n^{(eq)}_0}{Tg_2},
\label{omega}
\eeq
where $q$ stands for the wave-vector and $c_0^2= \rho_0(g_2- g^2_1n^{(eq)}_0/T)$ denotes speed of first sound. 

The value of $\tau$ can be related to the relaxation in Eq.(\ref{eq_dis}) as $ \tau^{-1} \sim \epsilon $. Thus,  at $T \to 0$, where $\tau \sim \exp(\Delta/T) \to \infty$,
and at $T\to \infty$, where $\tau \sim 1/\Delta(T) \to \infty$, the dissipation in Eq.(\ref{omega}) vanishes along with the dissipation of the shear modulus.

\section{Discussion and perspectives}\label{sum}
We have introduced the alternative to the traditionally accepted "boiling off" model of pinning gliding dislocations by \hee3 impurities: Quantum and thermal fluctuations of dislocation shape lead to decoupling of dislocations from the Peierls and \hee3-pinning potentials at temperatures determined by the energy of  kink-antikink pair. 

The analysis of the shape of the shear modulus vs temperature at significantly different \hee3 concentrations, Ref.\cite{Beamish}, and its comparison with the results of Monte Carlo simulations strongly indicate that the relaxation time for establishing equilibrium between \hee3 impurities bound to dislocations and those which are free in the bulk is very large on any experimental time scale. This conjecture is supported by the estimates of the values of the trapping potential for \hee3 and is consistent with the general nature of the narrow-band impuriton \cite{Andreev_uspekhi}. The formation of the two subsystems of \hee3 atoms---free in the bulk and bound to dislocations---is most likely to occur during \he4 crystal growth. We find this problem as very important for theoretical and experimental studies. 

It is shown that the shear modulus softening consists of two contributions: first, from thermally suppressed gap and, second, from dissipation. In general, these two effects are independent from each other so that the puzzle of "missing dissipation" can be resolved without envoking a wide distribution of relaxation times in relatively good quality crystals. 

The minimal model leading to the similarity between the TO NCRI response and the generalized shear modulus has been developed. It is shown that the strong suppression of the superfluidity along dislocation core by  dislocation kinks locks in both responses. In light of the recent work \cite{Beamish_Balibar} we note that the data which are well above the "red line" in Fig.1, Ref.\cite{Beamish_Balibar}, does not exclude the genuine NCRI anomaly whose similarity to the mechanical response is determined by the mechanism described above. 

The dissipation similarly seen in the TO and the shear modulus measurements  is proposed to be attributed to
the bath of dislocation kinks living along the dislocation network. The nature of such dissipation (its frequency and temperature dependencies) is linked to the fundamental question of how dissipation emerges in a network of 1d systems and, therefore, deserves close attention. Of a specific interest is studying the dissipation in the system of superclimbing dislocations, where the spectrum is not sound-like anymore \cite{sclimb}. 

Another interesting aspect is how the SF fraction and its excitations can be detected in the UMass-Sandwich-type setup \cite{Hallock}. Currently, the detected flow is d.c. and is in the overcritical (strongly non-linear) regime and, therefore, is not suitable for answering these questions. Thus, it is important to devise and implement the a.c. current sandwich-type setup in order to study the linear response of the SF network directly.

\begin{acknowledgements}
We are grateful to John Beamish for providing the experimental data and useful discussions.  We also thank Nikolay Prokof'ev and Boris Svistunov for many stimulating discussions. 
This work was supported by the  National Science Foundation
under Grant No.PHY1005527, and by a grant of computer time from the CUNY HPCC under NSF Grants CNS-0855217 and CNS - 0958379.

\end{acknowledgements}



\begin{thebibliography}{99}
\bibitem{note} A good fit can only be achieved for the very unrealistic values: $n_d \sim 10^{-2}, X_3 \sim 10^{-2}, E_a=0.2-0.3$K.
\end{thebibliography}
\end{document}